\documentclass[11pt]{article}
\usepackage[margin=1in]{geometry}
\usepackage{amsmath,amssymb,amsthm,mathtools}
\usepackage{microtype,xcolor}
\usepackage{tikz}
\usepackage{xurl}
\usepackage[numbers,sort&compress]{natbib}
\usepackage[colorlinks=true,linkcolor=blue!40!black,citecolor=blue!40!black,urlcolor=blue!40!black]{hyperref}
\newtheorem{theorem}{Theorem}[section]
\newtheorem{proposition}[theorem]{Proposition}
\newtheorem{lemma}[theorem]{Lemma}
\newtheorem{corollary}[theorem]{Corollary}
\theoremstyle{remark}

\newcommand{\F}{\mathbb F}
\newcommand{\R}{\operatorname{R}}
\newcommand{\aR}{\widetilde{\operatorname{R}}}
\newcommand{\one}{\mathbf1}

\newcommand{\GT}{\mathsf{GT}}
\newcommand{\WT}{\mathsf{WT}}
\DeclareMathOperator{\tr}{tr}

\hypersetup{pdfauthor={Emile Anand},pdftitle={The Asymptotic Rank of the Five-Cycle Tensor}}
\title{The Asymptotic Rank of the Five-Cycle Tensor}
\author{Emile Anand\\
\small Georgia Institute of Technology\\
\small\texttt{emile@gatech.edu}}
\date{}

\begin{document}
\maketitle

\begin{abstract}
Graph tensors generalize the matrix multiplication tensor, and their
asymptotic rank measures both the complexity of the associated
multilinear forms and the cost of creating networks of entangled
states. The five-cycle is the first graph beyond the triangle, and its
asymptotic rank was only known to lie between \(n^4\) and
\(n^{25/6}\). By Strassen's duality theorem, asymptotic rank is
controlled by spectral points, and each spectral point assigns prices
to the edges of a graph that behave like distances. These distances
always come from a CAT(0) space, which is a space of nonpositive
curvature in which every triangle is at least as thin as a Euclidean
triangle with the same side lengths. Euclidean spaces, trees and
hyperbolic spaces are all examples. This turns upper bounds on tensor
rank into questions about the geometry of a few points. We prove a
sharp inequality for five points in a CAT(0) space. If \(S\) is the sum
of the five cyclic side lengths, \(L\) is the sum of the five
diagonals, and \(r_i\) is the sum of the distances from the \(i\)th
point to the others, then \(L+r_{e+3}\le3S\) whenever \(\{e,e+1\}\) is
a longest side, and the constant \(3\) is best possible. Combined with
Zhou's copositive constraint on spectral points, this shows that \(L\)
is at most \(49/12\), which is sharp. Together with the exact exponent
of rectangular matrix multiplication, it gives
\(\aR_{\F}(\GT_n(C_5))=n^{49/12}\). We also give bounds for pentagons
with unequal bond dimensions and for graphs with fractional pentagon
packings, and we determine the exponent of separated decompositions of
\(\tr(A_0A_1A_2A_3A_4)\). As an application to quantum information, we
show that five parties who share GHZ states and use stochastic local
operations need asymptotically \(49/12\) GHZ states per copy to create
a pentagon of qubit EPR pairs.
\end{abstract}

\section{Introduction}

A graph tensor is built by placing an identity matrix on each edge of a
graph and grouping the two ends of every edge with the vertices they
meet. For the triangle this gives the matrix multiplication tensor, and
longer cycles give iterated matrix products. The rank of a graph tensor
is the number of fully separated products needed to write down the
associated multilinear form. Its asymptotic rank has a second
interpretation as well, since it measures how many GHZ states are needed
to produce the corresponding network of entangled pairs~\cite{CVZ}.

Importantly, the five-cycle is the first case that goes beyond matrix multiplication.
Let \(T_5(n)=\GT_n(C_5)\) be the five-cycle tensor with edge dimension
\(n\) and one tensor factor per vertex, and set
\[
\gamma_5=\log_2\aR_{\F}(T_5(2)).
\]
Here, flattenings show that \(\gamma_5\ge4\). Zhou's completely positive
framework combines a quadratic constraint on spectral points with the
triangle inequality and shows that \(\gamma_5\le25/6\)~\cite{Zhou}. The
value \(25/6\) is the true optimum of those constraints, so any
improvement has to use more information about which distance matrices
can come from spectral points.

The extra information we use is a five-point inequality. For five
labeled points \(z_0,\ldots,z_4\), with indices read modulo five, we
write
\begin{equation}\label{eq:metric-notation}
\begin{aligned}
s_i&=d(z_i,z_{i+1}),\\
b_i&=d(z_i,z_{i+2}).
\end{aligned}
\end{equation}
We call the \(s_i\) sides and the \(b_i\) diagonals, even though a
diagonal need not be longer than a side. The vertex opposite the side
\(\{e,e+1\}\) is \(e+3\).

\begin{theorem}[Five-point selection inequality]\label{thm:intro-geometry}
Let \(z_0,\ldots,z_4\) be points in a CAT(0) space, not necessarily
distinct. Let \(S=\sum_i s_i\) be the sum of the sides, let
\(L=\sum_i b_i\) be the sum of the diagonals, and let
\(r_i=\sum_j d(z_i,z_j)\) be the sum of the distances from \(z_i\) to
the other four points. If \(s_e=\max_j s_j\), then
\[
L+r_{e+3}\le3S.
\]
In particular \(L+\min_i r_i\le3S\). The constant \(3\) is best
possible, and this is already the case for points on the real line.
\end{theorem}

Notice that the theorem tells us which vertex to use. It is the vertex
opposite a longest side, and this is stronger than saying that some
vertex works. In the plane the inequality holds for every labeling of
the points, including labelings for which the closed polygonal path
crosses itself.

\begin{theorem}[Sharp copositive metric bound]\label{thm:intro-opt}
Let \(P_{ij}=d(z_i,z_j)\) for five points in a CAT(0) space, not
necessarily distinct, and suppose that \(J-2P\) is copositive. Then
\begin{equation}
\sum_{i=0}^4P_{i,i+2}\le\frac{49}{12}.
\end{equation}
Equality holds for a configuration of three distinct points in the
plane, and the bound remains sharp if copositivity is replaced by
positive semidefiniteness.
\end{theorem}

The extremal configuration in Theorem~\ref{thm:intro-opt} uses two of
its points twice, so repeated labels have to be allowed. We stress that
Theorem~\ref{thm:intro-opt} is a statement about metric geometry and
does not mention tensors at all.

To pass from geometry to tensors we use Strassen's duality
theorem~\cite{Strassen1988,WZ}, Zhou's quadratic
constraint~\cite{Zhou}, a curvature theorem for spectral wire
metrics~\cite{Geometry}, and an exact formula for the exponent of
rectangular matrix multiplication~\cite{Rectangle}. The precise
statements are given in Section~\ref{sec:spectral}.

\begin{theorem}[Five-cycle exponent]\label{thm:intro-rank}
For every integer \(n\ge2\),
\begin{equation}
\aR_{\F}(T_5(n))=n^{49/12}.
\end{equation}
Equivalently, \(\gamma_5=49/12\).
\end{theorem}

The upper bound comes from Theorem~\ref{thm:intro-opt} together with
the curvature and quadratic constraints. For the lower bound we group
the five tensor factors into three blocks, which turns \(T_5(n)\) into
rectangular matrix multiplication of shape \((n^3,n,n)\). By
Theorem~\ref{thm:rect} the exponent of this shape is \(49/12\), so the two
bounds agree.

\subsection{Outline of the proof}

By Reshetnyak's majorization theorem, a closed five-gon in a CAT(0)
space is majorized by a planar polygon with the same side lengths and
with distances between vertices that are at least as large~\cite{Reshetnyak}.
This means that it is enough to prove Theorem~\ref{thm:intro-geometry}
in the plane. There we scale a longest side to length one and fix the
directions of all five edges. The other four side lengths then range
over a box cut by two linear closure equations, and \(L+r_{e+3}-3S\) is
a convex function of these lengths, so we only need to check the
vertices of the resulting polytope. At a vertex, at least two of the
free lengths are equal to \(0\) or \(1\). If two of them are zero we are
left with a triangle with repeated labels, and if one is zero and one
is equal to one we are left with a quadrilateral with two longest
sides. Both of these cases have short geometric proofs. If two of them
are equal to one we get a pentagon with three longest sides. Up to
symmetry this case splits into four subcases, and each of them follows
from the triangle inequality together with Ptolemy's inequality, the
quadrilateral inequality, or the parallelogram law applied at a
suitable vertex.

Once we have Theorem~\ref{thm:intro-geometry}, the bound in
Theorem~\ref{thm:intro-opt} follows quickly. We pick a vertex \(i\)
with \(L+r_i\le3S\) and set \(v=3\one-e_i\). Then
\begin{align*}
48L&\le36(L+S)-12r_i\\
&=2v^\top Pv\\
&\le(\one^\top v)^2\\
&=196.
\end{align*}
The first step is the choice of \(i\), and the equality uses
\(P_{ii}=0\). The last inequality is copositivity of \(J-2P\) applied
to \(v\), which is nonnegative with entries summing to \(14\). Hence
\(L\le49/12\).

\subsection{Related work}

Christandl, Vrana and Zuiddam studied the asymptotic rank of graph
tensors and its connections with algebraic complexity, entanglement and
communication complexity~\cite{CVZ}. Christandl and Zuiddam treated odd
cycles as iterated matrix multiplication and gave rank constructions
for them~\cite{CZ}. The theory of the asymptotic spectrum of tensors
goes back to Strassen~\cite{Strassen1987,Strassen1988,Strassen1991}.
For background on tensor rank and matrix multiplication we refer to the
book of Bürgisser, Clausen and Shokrollahi~\cite{BCS}, and we follow the
presentation in the survey of Wigderson and Zuiddam~\cite{WZ}.
Christandl, Vrana and Zuiddam also constructed explicit families of
spectral points over the complex numbers, which are known as the
quantum functionals~\cite{CVZuniversal}.

There is a long history of upper bounds on the rectangular exponent,
starting with the method of Coppersmith and Winograd~\cite{CW} and
continuing with more recent refinements for rectangular
shapes~\cite{LGU,ADVXXZ}. The bound \(\omega\le9/4\) also has an
independent elementary proof~\cite{Pass}. Our lower bound uses the exact
rectangular formula of~\cite{Rectangle}, which determines the
exponent for every shape.

Zhou's quadratic constraint gives upper bounds for graph tensors inside
a completely positive optimization framework, and it yields the bound
\(m-1+1/(m+1)\) for odd cycles~\cite{Zhou}. What we add is a sharp
five-point constraint that fits into this framework. Majorization in
CAT(0) spaces is classical, and the new ingredients here are the
selection inequality and its proof. We do not claim that the
inequalities used in that proof characterize CAT(0) metrics or spectral
wire metrics.

Two of our applications are about restricted models, and it is worth
saying exactly how. A fully separated trace decomposition is weaker
than a general arithmetic circuit, because a circuit may reuse
intermediate matrix products. The quantum statement is about exact
conversion on a successful branch of a stochastic local operation, and
we give no bound on the probability of success.

\section{Preliminaries}\label{sec:prelim}

\subsection{Restriction, rank and graph tensors}

We fix a field \(\F\) and consider tensors of order at least two. For
tensors of the same order we write \(T\preceq U\) if \(T\) is the image
of \(U\) under a tensor product of local linear maps. The rank \(\R(T)\)
is the smallest number of simple tensors that sum to \(T\).
Equivalently, \(\R(T)\le r\) if and only if \(T\preceq\Delta_r\), where
\(\Delta_r=\sum_{a=1}^r e_a^{\otimes N}\) is the diagonal tensor of
order \(N\). The Kronecker product groups corresponding local factors
together. Since rank is submultiplicative, the limit
\begin{equation}
\aR(T)=\lim_{k\to\infty}\R(T^{\otimes k})^{1/k}
      =\inf_{k\ge1}\R(T^{\otimes k})^{1/k}
\end{equation}
exists. Restriction does not increase rank or asymptotic rank. Adding
or deleting a one-dimensional local factor changes neither quantity,
as one can see by inserting a nonzero vector and contracting with a
linear functional.

For distinct vertices \(i,j\) of a graph on \(N\) vertices, let
\(E_{ij}(q)\) be the tensor \(\sum_{a=1}^q e_a\otimes e_a\) placed at
\(i\) and \(j\), with one-dimensional factors at all other vertices.
Given edge dimensions \(q=(q_e)_{e\in E(G)}\) we set
\[
\WT_G(q)=\bigotimes_{e=\{i,j\}\in E(G)}E_{ij}(q_e),
\]
and we write \(\GT_n(G)\) when every edge has dimension \(n\). With the
usual cyclic order, the five-cycle tensor is
\begin{equation}\label{eq:tensor}
T_5(n)=\sum_{x_0,\ldots,x_4\in[n]}
e_{x_4,x_0}\otimes e_{x_0,x_1}\otimes e_{x_1,x_2}
\otimes e_{x_2,x_3}\otimes e_{x_3,x_4}.
\end{equation}
Each local factor has dimension \(n^2\), and
\(T_5(n)^{\otimes k}\simeq T_5(n^k)\).

\subsection{Spectral wire prices and the inputs we use}\label{sec:spectral}

Fix an order \(N\). A normalized spectral point \(\phi\) is a
nonnegative real function on tensors that is additive under direct
sums, multiplicative under Kronecker products and monotone under
restriction, and that satisfies \(\phi(\Delta_r)=r\). We write
\(\Phi_{N,\F}\) for the set of all such functions, which is called the
asymptotic spectrum. Strassen's duality theorem~\cite{Strassen1988}
(see also~\cite{WZ}) says that
\begin{equation}\label{eq:duality}
\aR_{\F}(T)=\max_{\phi\in\Phi_{N,\F}}\phi(T).
\end{equation}

For \(\phi\in\Phi_{N,\F}\) and \(i\ne j\) we define the wire prices
\[
P_{ij}^{(\phi)}=\log_2\phi(E_{ij}(2)),
\]
and we set \(P_{ii}^{(\phi)}=0\).
They are symmetric and lie in \([0,1]\), because a wire restricts to a
nonzero simple tensor and \(E_{ij}(2)\) is a restriction of
\(\Delta_2\). If we join two wires by a local contraction we get
\(E_{ij}(2)\preceq E_{ik}(2)\otimes E_{kj}(2)\), and hence
\(P_{ij}\le P_{ik}+P_{kj}\). The prices therefore form a pseudometric.

\begin{lemma}\label{lem:wire-base}
For every integer \(q\ge1\) we have \(\phi(E_{ij}(q))=q^{P_{ij}}\).
\end{lemma}

\begin{proof}
The case \(q=1\) is clear. For \(q\ge2\) let
\(a_k=\lfloor k\log_2q\rfloor\). Monotonicity in the dimension and
multiplicativity give
\[
\phi(E_{ij}(2))^{a_k}
\le\phi(E_{ij}(q))^k
\le\phi(E_{ij}(2))^{a_k+1}.
\]
Taking \(k\)th roots and letting \(k\to\infty\) proves the claim.
\end{proof}

If we combine the lemma with~\eqref{eq:duality}, we obtain
\begin{equation}\label{eq:weighted-spectral}
\log_2\aR_{\F}(\WT_G(q))
=\max_{\phi\in\Phi_{N,\F}}
\sum_{e=\{i,j\}\in E(G)}(\log_2q_e)\,P_{ij}^{(\phi)}.
\end{equation}
Our first input is a theorem of Zhou.

\begin{theorem}[{Zhou~\cite[Lemma~5.1]{Zhou}}]\label{thm:zhou}
For every \(N\), every field \(\F\), every \(\phi\in\Phi_{N,\F}\) and
every \(a\in\mathbb R_{\ge0}^N\),
\[
4\sum_{i<j}a_i a_jP_{ij}^{(\phi)}\le
\Bigl(\sum_i a_i\Bigr)^2.
\]
Equivalently, \(J-2P^{(\phi)}\) is copositive.
\end{theorem}

A symmetric matrix is copositive if its quadratic form is nonnegative
on the nonnegative orthant~\cite{Dur}. The theorem does not say that
\(J-2P^{(\phi)}\) is positive semidefinite. We will refer to
Theorem~\ref{thm:zhou} as the quadratic constraint (Q).

The other two inputs are the curvature theorem for spectral wire
metrics and the exact rectangular exponent.

\begin{theorem}[{\cite{Geometry}}]\label{thm:curvature}
For every spectral point over \(\F\), the wire-price pseudometric,
after points at distance zero are identified, embeds isometrically in a
CAT(0) space.
\end{theorem}

The curvature theorem in~\cite{Geometry} is stated for every spectral
space, and the construction there attaches a spectral space to every
spectral point. We need the statement for all spectral points, since a
result about some particular family of Euclidean spectral points would
not be enough for our argument.

For nonnegative integers \(a,b,c\) let
\(\omega_{\F}(a,b,c)=\log_2\aR_{\F}(\langle2^a,2^b,2^c\rangle)\) be the
exponent of rectangular matrix multiplication of shape
\((2^a,2^b,2^c)\). This agrees with the usual rectangular exponent, and
we will only need integer shapes.

\begin{theorem}[{\cite{Rectangle}}]\label{thm:rect}
For integers \(a\ge b\ge c>0\),
\[
\omega_{\F}(a,b,c)
=a+b+\frac{\max\{(a+b)c-ab,\,0\}^{2}}{4abc}.
\]
If one of the three numbers is zero, the exponent is the sum of the
other two.
\end{theorem}

In particular,
\begin{equation}\label{eq:omega311}
\omega_{\F}(3,1,1)=4+\frac1{12}=\frac{49}{12}.
\end{equation}
This value really does need the full rectangular formula, and it cannot
be obtained from a bound on the square exponent alone.

\subsection{CAT(0) spaces and majorization}

A CAT(0) space is a geodesic metric space in which every geodesic
triangle is no fatter than its comparison triangle in the Euclidean
plane~\cite{BH}. We will use Reshetnyak's majorization
theorem~\cite{ReshetnyakOrig} in the form given in~\cite{Reshetnyak}.

\begin{theorem}[Polygonal majorization]\label{thm:majorization}
Let \(z_0,\ldots,z_{m-1}\) be the vertices of a closed geodesic polygon
in a CAT(0) space. There are points \(v_0,\ldots,v_{m-1}\), in this
cyclic order on the boundary of a convex region in the Euclidean plane,
such that
\begin{align*}
|v_i-v_{i+1}|&=d(z_i,z_{i+1}),\\
|v_i-v_j|&\ge d(z_i,z_j)
\end{align*}
for all \(i,j\), with indices taken modulo \(m\). We allow degenerate polygons and repeated vertices.
\end{theorem}

\section{Proof of Theorem~\ref{thm:intro-geometry}}
\label{sec:fivepoint-geometry}

In this section \(z_0,\ldots,z_4\) are five points, not necessarily
distinct, and we use the notation of~\eqref{eq:metric-notation} and
Theorem~\ref{thm:intro-geometry} together
with \(d_{ij}=d(z_i,z_j)\). All indices are taken modulo five. The proof
has three steps. First we reduce to the plane, then we reduce further
to finitely many boundary patterns, and finally we treat each pattern
separately. The planar argument never uses convexity of the polygon.

\begin{lemma}[Planar case]\label{lem:planar-selection}
Let \(v_0,\ldots,v_4\in\mathbb R^2\). If
\(|v_0-v_1|=\max_j|v_j-v_{j+1}|\), then \(L+r_3\le 3S\).
\end{lemma}

\begin{proof}[Proof of Theorem~\ref{thm:intro-geometry}, assuming Lemma~\ref{lem:planar-selection}]
After relabeling we may assume that \(e=0\). We join consecutive points
by geodesics and apply Theorem~\ref{thm:majorization}. The majorizing
map does not increase distances and preserves arclength along the
boundary. Each marked boundary arc is a straight segment, because its
length equals \(d(z_i,z_{i+1})\), while its chord is at least this
distance and at most the length of the arc. Hence the planar points
satisfy
\begin{align*}
|v_i-v_{i+1}|&=s_i,\\
|v_i-v_j|&\ge d_{ij}.
\end{align*}
In particular \(v_0v_1\) is still a longest side. If we write
\(\widehat S,\widehat L,\widehat r_3\) for the planar quantities, then
\(\widehat S=S\), \(\widehat L\ge L\) and \(\widehat r_3\ge r_3\), and
Lemma~\ref{lem:planar-selection} gives
\[
L+r_3\le\widehat L+\widehat r_3\le3\widehat S=3S.
\]
Degenerate majorizing polygons and sides of length zero cause no
difficulty, because the planar lemma allows them.

To see that the constant \(3\) is sharp, place the points
\((z_0,\ldots,z_4)=(0,0,2,2,1)\) on the real line. Then \(S=4\) and
\(L=8\), the vertex sums are \((5,5,5,5,4)\), and
\(L+\min_i r_i=12=3S\).
\end{proof}

\subsection{Reduction to boundary patterns}
\label{subsec:box-section}

\begin{lemma}\label{lem:boundary-reduction}
To prove Lemma~\ref{lem:planar-selection} it is enough to consider
configurations with \(s_0=1\) in which at least two of the lengths
\(s_1,s_2,s_3,s_4\) lie in \(\{0,1\}\).
\end{lemma}

\begin{proof}
If all five points coincide there is nothing to prove. Otherwise we
scale so that \(s_0=1\), and we fix unit vectors \(u_0,\ldots,u_4\)
pointing along the edges, choosing an arbitrary direction for any edge
of length zero. We let the other four lengths vary over the compact
polytope
\[
\mathcal B=\Bigl\{t\in[0,1]^4\Bigm|u_0+\sum_{j=1}^4t_j u_j=0\Bigr\}.
\]
With \(t_0=1\), \(v_0(t)=0\) and \(v_k(t)=\sum_{j=0}^{k-1}t_j u_j\), the
closure condition says that \(v_5(t)=v_0(t)\). Each distance
\(|v_i(t)-v_j(t)|\) is the norm of an affine function of \(t\), so it is
convex in \(t\). As a result the function
\[
F(t)=L(t)+r_3(t)-3\sum_{j=0}^4t_j
\]
is convex on \(\mathcal B\), and it suffices to show that \(F\le0\) at
the vertices of \(\mathcal B\).

At a vertex of \(\mathcal B\), at most two coordinates lie strictly
between \(0\) and \(1\). Indeed, if three of them did, the
corresponding direction vectors in \(\mathbb R^2\) would satisfy a
nontrivial linear relation, and moving along this relation in either
direction would keep us inside \(\mathcal B\). The same argument works
when the closure equations have rank less than two.
\end{proof}

So at a vertex of \(\mathcal B\) there are two zero edges, or one zero
edge and one more unit edge, or two more unit edges. We treat these
three cases in turn. In each case the letters \(a,b,c,d,p,q\) denote
lengths that are local to that case, and they should not be confused
with the diagonals \(b_i\).

\subsection{Two zero edges}

In this case the points form at most three clusters of consecutive
equal labels. When there are three clusters, their cyclic
multiplicities are \((3,1,1)\) or \((2,2,1)\). We call the positions of
the clusters \(X,Y,Z\) and set \(a=XY\), \(b=YZ\) and \(c=ZX\).

If the multiplicities are \((3,1,1)\), then the tripled point \(X\) is
opposite every nonzero side. A direct count gives \(S=a+b+c\),
\(L=2a+2c\) and \(r_X=a+c\), so
\begin{align*}
L+r_X-3S&=(2a+2c)+(a+c)-3(a+b+c)\\
&=-3b\\
&\le0.
\end{align*}
If the multiplicities are \((2,2,1)\), we take the order \(X,X,Y,Y,Z\),
so that \(L=3a+b+c\). When \(a\) is a longest side the opposite label
is \(Z\), and \(L+r_Z=3S\). When \(b\) is a longest side the opposite
label is \(X\), and \(L+r_X-3S=2a-2b-c\le0\). When \(c\) is a longest
side the same computation gives \(2a-b-2c\le0\). Further coincidences
among the points are covered by letting some of \(a,b,c\) vanish.

\subsection{One zero edge and two longest sides}

A reflection that fixes the distinguished side and its opposite vertex
reduces this case to two configurations. For four points \(A,B,C,D\)
in the plane we write \(a=AB\), \(b=BC\), \(c=CD\) and \(d=DA\) for the
sides and \(p=AC\) and \(q=BD\) for the diagonals. We assume that
\(a\ge b,c,d\) and that at least one of \(b,c,d\) is equal to \(a\).

For the labels \(A,B,C,C,D\) the selected vertex is \(C\), and the
triangle inequality is all we need, since
\begin{align*}
L+r_C-3S&=3p+q-3a-b-c-3d\\
&\le3(c+d)+(b+c)-3a-b-c-3d\\
&=3(c-a)\\
&\le0.
\end{align*}

For the labels \(A,B,B,C,D\) the selected vertex is again \(C\), and
\[
L+r_C-3S=2(p+q-a-c)-3d.
\]
We therefore have to show that \(p+q\le a+c+\tfrac32d\), and we split
the argument into three subcases.

Suppose first that \(b=a\). We place \(A=(0,0)\), \(C=(p,0)\),
\(B=(p/2,h)\) and \(D=(x,t)\). If \(0\le x\le p\), then \(q\le a+|t|\)
and \(p-c\le x\), so
\[
p+q-a-c\le x+|t|\le\sqrt2\,d\le\tfrac32d.
\]
If \(x<0\) we use \(p\le c\) and \(q\le a+d\), and if \(x>p\) we use
\(p\le d\) and \(q\le a+c\). In both cases we get the stronger bound
\(p+q-a-c\le d\).

Suppose next that \(c=a\). The identity
\[
p^2+q^2-a^2-c^2=2(A-D)\cdot(B-C)\le2db\le2ad
\]
gives \(p+q\le2\sqrt{a^2+ad}\le2a+d\).

Finally, suppose that \(d=a\). The quadrilateral inequality
\(p^2+q^2\le a^2+b^2+c^2+d^2\) gives
\[
p+q\le\sqrt{2(p^2+q^2)}
\le\sqrt{6a^2+2c^2}\le\tfrac52a+c,
\]
where the last step can be checked by squaring, since
\(a^2/4+5ac-c^2\ge0\) for \(0\le c\le a\).

\subsection{Three longest sides}

We scale the common longest length to one and write \(A,B,C,D,E\) for
\(v_0,\ldots,v_4\). Up to rotation and reflection, the side lengths
read in cyclic order are either \((1,1,1,x,y)\) or \((1,1,x,1,y)\) with
\(0\le x,y\le1\), and we have to prove \(L+r_{e+3}\le3S\) for each unit side \(e\).

Apart from the triangle inequality, we use three facts about points in
the plane. The first is Ptolemy's inequality
\(PR\cdot QT\le PQ\cdot RT+PT\cdot QR\), which holds for any four
points. The second is the quadrilateral inequality
\(PR^2+QT^2\le PQ^2+QR^2+RT^2+TP^2\), which is true because the
difference between the two sides equals \(|P-Q+R-T|^2\). The third is
the parallelogram law \(OP^2+OQ^2=\tfrac12\bigl(|P+Q-2O|^2+PQ^2\bigr)\).
Together with \((OP+OQ)^2\le2(OP^2+OQ^2)\) it gives
\begin{equation}\label{eq:par}
OP+OQ\le\sqrt{|P+Q-2O|^2+PQ^2}.
\end{equation}
We also need two elementary estimates. If \(0\le p,q\le P\) and
\(P>0\), then \((P-p)(P-q)\ge0\), which can be rewritten as
\begin{equation}\label{eq:box}
p+q\le P+\frac{pq}{P}.
\end{equation}
For \(a,b\ge0\) we have
\(\sqrt{(a+b)^2+1}\le\sqrt{a^2+1}+b\), and for \(a>0\) we also have
\(\sqrt{a^2+1}\le a+\frac1{2a}\).

\subsubsection*{The pattern \((1,1,x,1,y)\)}

Here \(AB=BC=DE=1\), \(CD=x\) and \(EA=y\), and we set \(s=AC\). The
unit sides are \(0\), \(1\) and \(3\), and their opposite vertices are
\(D\), \(E\) and \(B\). The reflection \(i\mapsto2-i\pmod5\) swaps
sides \(0\) and \(1\) and also swaps \(x\) and \(y\), so it is enough to
deal with the vertices \(B\) and \(D\). Since
\(D+E-2B=(A-B)+(C-B)+(D-C)+(E-A)\) and \(|(A-B)+(C-B)|^2=4-s^2\), the
bound~\eqref{eq:par} at \(B\) together with \(DE=1\) gives
\begin{equation}\label{eq:parB}
BD+BE\le\sqrt{\bigl(\sqrt{4-s^2}+x+y\bigr)^2+1}\le\sqrt{5-s^2}+x+y.
\end{equation}

We begin with the vertex \(B\). The inequality \(L+r_B\le3S\) that we
need reads
\[
s+2(BD+BE)+CE+DA\le7+3x+3y.
\]
We estimate the left side by
\begin{align*}
s+2(BD+BE)+CE+DA
&\le s+2\sqrt{5-s^2}+2x+2y+CE+DA\\
&\le s+2\sqrt{5-s^2}+2\min(1,s)+3x+3y\\
&\le5+2+3x+3y\\
&=7+3x+3y.
\end{align*}
The first step is~\eqref{eq:parB}. The second uses the triangle
inequality, which gives \(CE\le\min(1+x,\,s+y)\) and
\(DA\le\min(1+y,\,s+x)\), and hence \(CE+DA\le x+y+2\min(1,s)\). The
third uses \(\min(1,s)\le1\) and the Cauchy-Schwarz bound
\(s+2\sqrt{5-s^2}\le\sqrt5\,\sqrt{s^2+(5-s^2)}=5\). Equality holds for
the equilateral configuration with \(x=y=0\) and \(s=1\).

Next we consider the vertex \(D\). Here we need
\[
s+2BD+CE+2DA+EB\le8+2x+3y.
\]
We first show that
\[
BD+CE\le1+x+\frac{1+x+xy}{1+x}\le2+x+\frac y2.
\]
This is~\eqref{eq:box} with \(P=1+x\). It applies because \(BD\) and
\(CE\) are both at most \(1+x\), and Ptolemy's inequality for
\(B,C,D,E\) gives \(BD\cdot CE\le BC\cdot DE+CD\cdot BE\le1+x(1+y)\).
The second step uses \(x/(1+x)\le1/2\), which holds because \(x\le1\). Writing the left side as
\(s+(BD+CE)+(BD+EB)+2DA\), we now get
\begin{align*}
s+2BD+CE+2DA+EB
&\le s+\Bigl(2+x+\frac y2\Bigr)+\bigl(\sqrt{5-s^2}+x+y\bigr)+2(1+y)\\
&=s+\sqrt{5-s^2}+4+2x+\tfrac72y\\
&\le\sqrt{10}+\tfrac12+4+2x+3y\\
&<8+2x+3y.
\end{align*}
The first step combines the bound just proved with~\eqref{eq:parB} and
\(DA\le1+y\). The third step uses the Cauchy-Schwarz bound
\(s+\sqrt{5-s^2}\le\sqrt{10}\) and \(y\le1\), and the last uses
\(\sqrt{10}+\tfrac12<4\).

\subsubsection*{The pattern \((1,1,1,x,y)\)}

Here \(AB=BC=CD=1\), \(DE=x\) and \(EA=y\). We set \(p=AC\), \(q=BD\)
and \(\delta=DA\). The unit sides are \(0\), \(1\) and \(2\), and their
opposite vertices are \(D\), \(E\) and \(A\). The reflection
\(i\mapsto3-i\pmod5\) swaps sides \(0\) and \(2\) and also swaps \(x\)
and \(y\), so it is enough to deal with \(D\) and \(E\). Note that
\(p,q\le2\) and \(\delta\le x+y\). Ptolemy's inequality and the
quadrilateral inequality for \(A,B,C,D\) give
\begin{equation}\label{eq:ABCD}
\begin{aligned}
pq&\le1+\delta,\\
p^2+q^2&\le3+\delta^2.
\end{aligned}
\end{equation}
For the vertex \(E\) we write
\(B+C-2E=(B-A)+(C-D)+(A-E)+(D-E)\). The first two vectors have length
one, and by the parallelogram law
\(|(A-E)+(D-E)|^2=2x^2+2y^2-\delta^2\). Let \(J\) be the square root of
this quantity. Applying~\eqref{eq:par} at \(E\) with \(BC=1\) gives
\begin{equation}\label{eq:parE}
CE+EB\le\sqrt{(2+J)^2+1}\le\frac94+J.
\end{equation}

We start with the vertex \(E\), where we need
\[
p+q+\delta+2(CE+EB)\le9+2x+2y.
\]
Adding twice the first inequality of~\eqref{eq:ABCD} to the second
gives
\[
(p+q)^2\le3+\delta^2+2(1+\delta)=(\delta+1)^2+4.
\]
Suppose first that \(\delta\le5/3\). Then
\begin{align*}
p+q+\delta+2(CE+EB)
&\le\sqrt{(\delta+1)^2+4}+\delta+4+2x+2y\\
&\le9+2x+2y.
\end{align*}
The first step uses the bound on \((p+q)^2\) together with
\(CE\le1+x\) and \(EB\le1+y\). For the second step, squaring shows that
\(\sqrt{(\delta+1)^2+4}+\delta\le5\) is equivalent to \(12\delta\le20\).

Suppose now that \(\delta>5/3\). Then
\begin{align*}
p+q+\delta+2(CE+EB)
&\le\Bigl(\delta+\frac74\Bigr)+\delta+2\Bigl(\frac94+J\Bigr)\\
&<2\delta+\frac74+\frac92+\frac{20}9\\
&<2\delta+9\\
&\le9+2x+2y.
\end{align*}
The first step uses~\eqref{eq:parE} and the bound
\(p+q\le(\delta+1)+\frac2{\delta+1}\le\delta+\frac74\), which follows
from the bound on \((p+q)^2\) and \(\delta>5/3\). The second step uses
\(J<\frac{10}9\), which holds because
\(J^2=2x^2+2y^2-\delta^2\le4-\delta^2<\frac{11}9\). The last step uses
\(\delta\le x+y\).

It remains to treat the vertex \(D\), where we need
\[
p+2q+2\delta+CE+EB\le8+2x+3y.
\]
Suppose first that \(\delta\le\sqrt2\). Then
\begin{align*}
p+2q+2\delta+CE+EB&\le5+2\delta+(1+x)+(1+y)\\
&\le7+3x+3y\\
&\le8+2x+3y.
\end{align*}
The first step uses \(CE\le1+x\), \(EB\le1+y\) and
\(p+2q\le\sqrt{5(p^2+q^2)}\le\sqrt{5(3+\delta^2)}\le5\), which follows
from the Cauchy-Schwarz inequality and~\eqref{eq:ABCD}. The second step
uses \(\delta\le x+y\), and the third uses \(x\le1\).

Suppose now that \(\delta>\sqrt2\). Then
\begin{align*}
p+2q+2\delta+CE+EB
&\le4+\sqrt{\delta^2-1}+2\delta+\frac94+J\\
&\le4+\frac94+\frac74+y+2\delta\\
&\le8+2x+3y.
\end{align*}
The first step uses~\eqref{eq:parE} and \(p+2q\le4+\sqrt{\delta^2-1}\).
To see the latter, note that \(p\le\sqrt{3+\delta^2-q^2}\)
by~\eqref{eq:ABCD}, that the function
\(q\mapsto2q+\sqrt{3+\delta^2-q^2}\) is increasing on \([0,2]\) because
\(3+\delta^2\ge5\), and that \(q\le2\). The last step uses
\(\delta\le x+y\). The second step uses the estimate
\[
\sqrt{\delta^2-1}+J\le\sqrt{2(\delta^2-1+J^2)}
=\sqrt{2(2x^2+2y^2-1)}\le\sqrt{2+4y^2}\le\frac74+y.
\]
Here the first inequality is the Cauchy-Schwarz inequality, the second
uses \(x\le1\), and the last holds because
\(3y^2-\frac72y-\frac{17}{16}\le0\) for \(0\le y\le1\).

This completes the boundary cases, and with them the proofs of
Lemma~\ref{lem:planar-selection} and Theorem~\ref{thm:intro-geometry}.

\section{Proof of Theorem~\ref{thm:intro-opt}}
\label{sec:metric-optimization}

Let \(P\) be the distance matrix of five points in a CAT(0) space such
that \(J-2P\) is copositive. By Theorem~\ref{thm:intro-geometry} there
is an index \(i\) with \(L+r_i\le3S\). We set \(v=3\one-e_i\). Then
\begin{align*}
48L&\le36(S+L)-12r_i\\
&=2v^\top Pv\\
&\le\Bigl(\sum_jv_j\Bigr)^2\\
&=196.
\end{align*}
The first step uses \(L+r_i\le3S\). The equality holds because
\(P_{ii}=0\), which gives \(v^\top Pv=18(S+L)-6r_i\). The last
inequality is copositivity of \(J-2P\) applied to the nonnegative
vector \(v\), whose entries sum to \(14\). This shows that
\(L\le49/12\).

To see that equality can occur, take the planar points in the order
\((A,A,B,B,C)\), where
\begin{align*}
A&=\Bigl(-\frac{35}{72},0\Bigr),\\
B&=\Bigl(\frac{35}{72},0\Bigr),\\
C&=\Bigl(0,\frac{7\sqrt{11}}{72}\Bigr).
\end{align*}
The nonzero distances are \(a=AB=35/36\) and \(b=AC=BC=7/12\). Then
\begin{align*}
L&=3a+2b=\frac{49}{12},\\
S&=a+2b=\frac{77}{36},\\
r_C&=4b=\frac73,
\end{align*}
so we also have \(L+r_C=3S\). For \(w\in\mathbb R^5\) let \(X=w_0+w_1\),
\(Y=w_2+w_3\) and \(Z=w_4\). Then
\[
w^\top(J-2P)w
=\frac{35}{36}(X-Y)^2+\frac1{36}(X+Y-6Z)^2\ge0,
\]
so \(J-2P\) is positive semidefinite and in particular copositive. This
proves both parts of Theorem~\ref{thm:intro-opt}. \qed

\medskip
This argument does not use Theorem~\ref{thm:curvature}. By that
theorem and Theorem~\ref{thm:zhou}, every spectral price matrix
satisfies the hypotheses of Theorem~\ref{thm:intro-opt}, so
Theorem~\ref{thm:intro-opt} gives the exact optimum of a relaxation of
the set of spectral price matrices. We do not know whether the
extremal configuration is the price matrix of an actual spectral point.

\begin{figure}[htbp]
\centering
\begin{tikzpicture}[scale=2.2,every node/.style={font=\small}]
\coordinate (a) at (-.486111,0);
\coordinate (b) at (.486111,0);
\coordinate (c) at (0,.32248);
\draw[thick] (a)--node[below]{\(35/36\)}(b);
\draw[thick] (a)--node[above left]{\(7/12\)}(c);
\draw[thick] (b)--node[above right]{\(7/12\)}(c);
\fill (a) circle (.025) node[left]{\(A=z_0=z_1\)};
\fill (b) circle (.025) node[right]{\(B=z_2=z_3\)};
\fill (c) circle (.025) node[above=3pt]{\(C=z_4\)};
\end{tikzpicture}
\caption{The extremal configuration for Theorem~\ref{thm:intro-opt}.
The five labels sit on three points of the plane, and the matrix
\(J-2P\) is positive semidefinite.}
\label{fig:optimizer}
\end{figure}
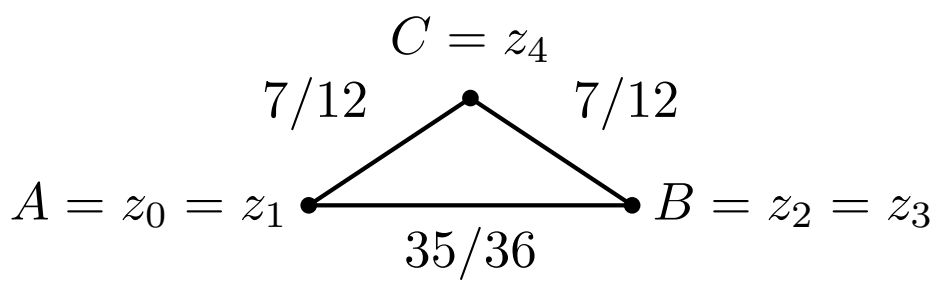

\section{Proof of Theorem~\ref{thm:intro-rank}}\label{sec:tensor-exponent}

All ranks in this section are taken over the fixed field \(\F\). By the
identity \(T_5(n)^{\otimes k}\simeq T_5(n^k)\) and
Proposition~\ref{prop:all-dimensions} below, it is enough to prove that
\(\gamma_5=49/12\).

\subsection{The upper bound}

Let \(\phi\) be a five-party spectral point with price matrix \(P\). We
relabel the vertices so that the edges of the cycle are the pairs
\(\{i,i+2\}\). By multiplicativity,
\[
\log_2\phi(T_5(2))=\sum_iP_{i,i+2}.
\]
By Theorem~\ref{thm:zhou} the matrix \(J-2P\) is copositive, and by
Theorem~\ref{thm:curvature} the matrix \(P\) is the distance matrix of five points
in a CAT(0) space. Theorem~\ref{thm:intro-opt} now gives
\(\log_2\phi(T_5(2))\le49/12\), and taking the maximum over \(\phi\)
in~\eqref{eq:duality} gives \(\gamma_5\le49/12\). It is worth pointing
out that copositivity is only applied to the original price matrix, so
we never need it to survive the passage to a planar majorizing polygon.

\subsection{The lower bound}

We return to the cyclic order of~\eqref{eq:tensor} and group the local
factors into the three blocks \(A=\{0,2\}\), \(B=\{1,3\}\) and
\(C=\{4\}\).
Three edges join \(A\) to \(B\), and one edge joins \(C\) to each of
the other two blocks. With \(u=(x_0,x_1,x_2)\), \(v=x_3\) and
\(w=x_4\), and after reordering the coordinates inside each block, the
grouped tensor is
\[
\sum_{u\in[n]^3,\ v,w\in[n]}
e_{u,w}\otimes e_{u,v}\otimes e_{v,w},
\]
which is rectangular matrix multiplication with edge dimensions
\((n^3,n,n)\). A simple five-party tensor remains simple after
grouping, so for every \(k\) we have
\[
\R_5(T_5(2)^{\otimes k})
\ge \R_3\bigl(\langle2^{3k},2^k,2^k\rangle\bigr).
\]
Taking logarithms, dividing by \(k\) and applying Theorem~\ref{thm:rect}, we
get
\[
\gamma_5\ge\omega_{\F}(3,1,1)
=3+1+\frac{((3+1)-3)^2}{4\cdot3}
=\frac{49}{12}.
\]
This completes the proof of Theorem~\ref{thm:intro-rank}. \qed

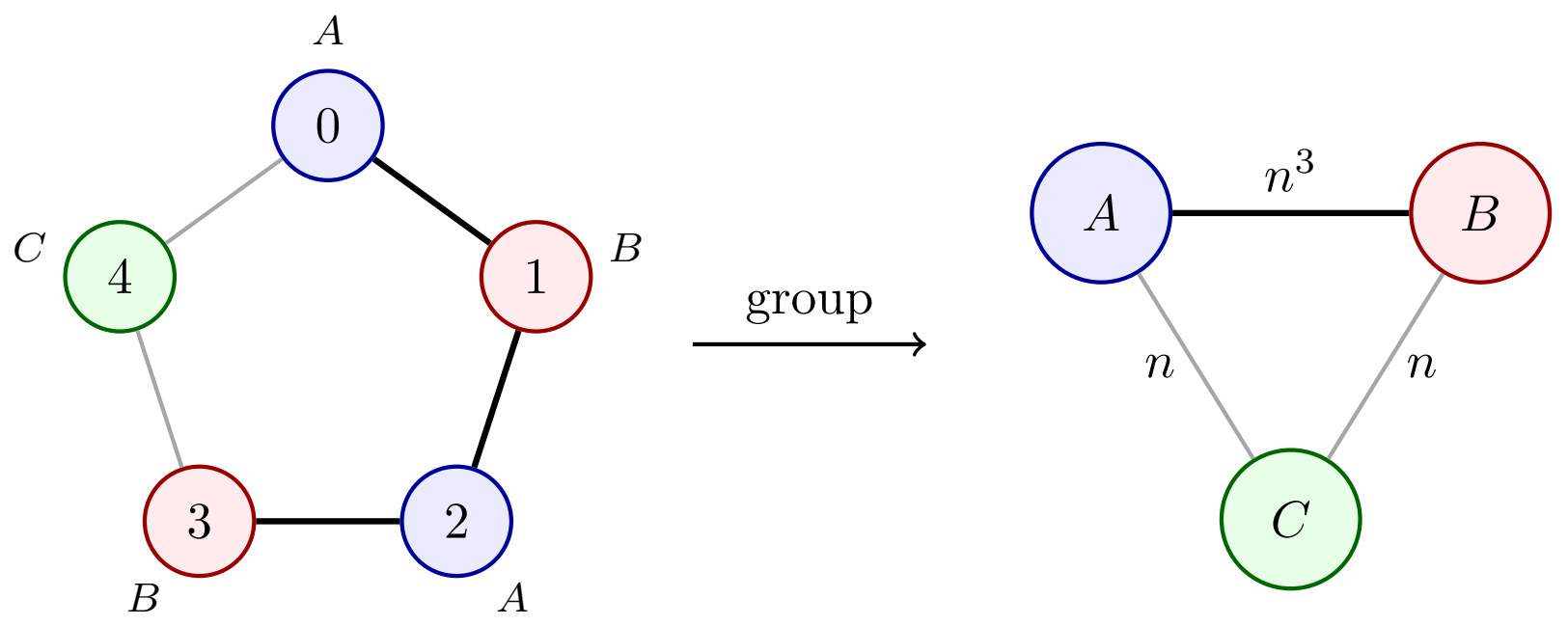
\begin{figure}[htbp]
\centering
\begin{tikzpicture}[
  every node/.style={font=\small},
  vtx/.style={circle,thick,minimum size=7.5mm,inner sep=0pt},
  big/.style={circle,thick,minimum size=9.5mm,inner sep=0pt},
  inA/.style={draw=blue!60!black,fill=blue!8},
  inB/.style={draw=red!60!black,fill=red!8},
  inC/.style={draw=green!40!black,fill=green!10},
  ab/.style={very thick,black},
  cx/.style={thick,gray!70}]
\foreach \i/\blk/\lab in {0/inA/A,1/inB/B,2/inA/A,3/inB/B,4/inC/C}{
  \node[vtx,\blk] (p\i) at ({90-72*\i}:1.5cm) {\(\i\)};
  \node[font=\scriptsize] at ({90-72*\i}:2.15cm) {\(\lab\)};}
\draw[ab] (p0)--(p1) (p1)--(p2) (p2)--(p3);
\draw[cx] (p3)--(p4) (p4)--(p0);
\draw[->,thick] (2.5,0)--(4.1,0) node[midway,above]{group};
\node[big,inA] (a) at (5.3,0.9) {\(A\)};
\node[big,inB] (b) at (7.9,0.9) {\(B\)};
\node[big,inC] (c) at (6.6,-1.2) {\(C\)};
\draw[ab] (a)--node[above]{\(n^3\)}(b);
\draw[cx] (a)--node[left,text=black]{\(n\)}(c);
\draw[cx] (b)--node[right,text=black]{\(n\)}(c);
\end{tikzpicture}
\caption{The grouping used for the lower bound, with blocks
\(A=\{0,2\}\), \(B=\{1,3\}\) and \(C=\{4\}\). The three dark edges of
the five-cycle all run between \(A\) and \(B\), and they merge into a
single index of dimension \(n^3\). The two gray edges join \(C\) to the
other two blocks.}
\label{fig:grouping}
\end{figure}

No other grouping of this kind does better, as the next proposition
shows.

\begin{proposition}\label{prop:groupings}
Among all ways of grouping the five vertices into three nonempty
blocks, the largest lower bound that Theorem~\ref{thm:rect} gives is
\(49/12\).
\end{proposition}

\begin{proof}
An edge inside a block contributes a nonzero local vector and does not
change the rank of the grouped tensor. The edges between blocks form a
connected Eulerian multigraph on three vertices with at most five
edges. If its edge multiplicities are \(a,b,c\), then \(a+b\), \(b+c\)
and \(c+a\) are even, so \(a,b,c\) all have the same parity. Up to
permutation the only possibilities are \((1,1,1)\), \((2,2,0)\) and
\((3,1,1)\), and their exponents are \(9/4\), \(4\) and \(49/12\).
\end{proof}

We should point out that Theorem~\ref{thm:intro-rank} is a statement
about the asymptotic exponent. It says nothing about the rank or the
border rank of the single tensor \(T_5(2)\), and the proof of the upper
bound does not produce an explicit decomposition.

\section{Applications}\label{sec:applications}

The upper bounds in
Sections~\ref{subsec:unequal} and~\ref{subsec:packing} use duality
together with Theorems~\ref{thm:zhou} and~\ref{thm:curvature}, and the
lower bounds use Theorem~\ref{thm:rect}.

\subsection{All bond dimensions and the trace of five matrices}

\begin{proposition}\label{prop:all-dimensions}
For every integer \(n\ge2\),
\begin{equation}\label{eq:all-dimensions}
\aR_5(T_5(n))=n^{49/12},
\end{equation}
and as \(n\to\infty\),
\begin{equation}\label{eq:ordinary-rank-exponent}
\R_5(T_5(n))=n^{49/12+o(1)}.
\end{equation}
\end{proposition}

\begin{proof}
Pairing edge indices gives \(T_5(a)\otimes T_5(b)\simeq T_5(ab)\), and
coordinate projections show that \(T_5(a)\preceq T_5(b)\) whenever
\(a\le b\). Write \(\gamma=49/12\). By the definition of asymptotic
rank, \(\log_2\R_5(T_5(2^m))=m\gamma+o(m)\), and sandwiching \(n\)
between consecutive powers of two proves~\eqref{eq:ordinary-rank-exponent}.
For fixed \(n\) we apply this along the sequence \(n^k\) and use
\(T_5(n)^{\otimes k}\simeq T_5(n^k)\), which gives
\[
\log\aR_5(T_5(n))
=\lim_{k\to\infty}\frac{\log\R_5(T_5(n^k))}{k}
=\gamma\log n. \qedhere
\]
\end{proof}

The connection between odd cycles and iterated matrix multiplication is
standard~\cite{CZ}. To make precise what the exponent measures, let
\(\sigma_5(n)\) be the smallest \(r\) for which there is an identity
\begin{equation}\label{eq:separated-trace}
\tr(A_0A_1A_2A_3A_4)
=\sum_{a=1}^{r}\prod_{i=0}^{4}\lambda_{a,i}(A_i),
\end{equation}
where each \(\lambda_{a,i}\) is a linear form in the entries of the
single matrix \(A_i\in\F^{n\times n}\).

\begin{corollary}\label{cor:trace}
We have \(\sigma_5(n)=n^{49/12+o(1)}\).
\end{corollary}

\begin{proof}
Expanding the trace gives
\[
\sum_{x_0,\ldots,x_4\in[n]}
(A_0)_{x_4,x_0}(A_1)_{x_0,x_1}(A_2)_{x_1,x_2}
(A_3)_{x_2,x_3}(A_4)_{x_3,x_4},
\]
and the coefficient tensor of this form is \(T_5(n)\). Each product on
the right of~\eqref{eq:separated-trace} is a simple tensor, and every
simple tensor arises in this way, so \(\sigma_5(n)=\R_5(T_5(n))\). The
claim now follows from Proposition~\ref{prop:all-dimensions}.
\end{proof}

This result is only about separated set-multilinear decompositions. A
general arithmetic circuit can multiply the matrices one after another
and reuse intermediate results, and \(\sigma_5(n)\) gives no lower
bound in that model. It also does not account for the cost of
evaluating the linear forms.

Iterated matrix multiplication is one of the central polynomials of
algebraic complexity theory. Nisan and Wigderson introduced the method
of partial derivatives for set-multilinear models~\cite{NW}, and lower
bounds for set-multilinear circuits computing iterated matrix
multiplication are the starting point of the superpolynomial lower
bounds for constant-depth circuits of Limaye, Srinivasan and
Tavenas~\cite{LST}. Those results concern many factors and circuits of
bounded depth, and they give bounds that are far from tight for any
fixed number of factors. Corollary~\ref{cor:trace} is of a different
kind. It fixes five factors and the simplest depth-two model, a sum of
products of linear forms, and it determines the exponent exactly.
Before this work, the best bounds on this exponent were \(4\) and
\(25/6\)~\cite{Zhou}.

\subsection{Pentagons with unequal bond dimensions}\label{subsec:unequal}

For integers \(q_0,\ldots,q_4\ge1\) let \(T_5(\mathbf q)\) be the cycle
tensor in which the edge \(\{i,i+1\}\) has dimension \(q_i\), and let
\(q_*=\min_i q_i\).

\begin{proposition}\label{prop:weighted-pentagon}
The asymptotic rank of \(T_5(\mathbf q)\) satisfies
\begin{equation}\label{eq:weighted-pentagon}
\max\Bigl\{q_*^{49/12},\ \frac{\prod_iq_i}{q_*}\Bigr\}
\ \le\ \aR_5(T_5(\mathbf q))
\ \le\ \frac{\prod_iq_i}{q_*^{11/12}}.
\end{equation}
The two bounds agree when all five dimensions are equal.
\end{proposition}

\begin{proof}
Fix a spectral point \(\phi\) and let
\(p_i=\log_2\phi(E_{i,i+1}(2))\in[0,1]\). By
Lemma~\ref{lem:wire-base}, \(\phi(E_{i,i+1}(q))=q^{p_i}\) for every
integer \(q\ge1\). Set \(w_i=\log_2q_i\) and \(t=\min_iw_i\). The upper
bound in Theorem~\ref{thm:intro-rank} gives \(\sum_ip_i\le49/12\), and
so
\[
\log_2\phi(T_5(\mathbf q))
=t\sum_i p_i+\sum_i(w_i-t)p_i
\le\frac{49}{12}t+\sum_i(w_i-t)
=\sum_iw_i-\frac{11}{12}t.
\]
Maximizing over \(\phi\) gives the upper bound. For the first lower
bound we project every edge down to dimension \(q_*\). For the second
we cut the cycle so that every edge except one of dimension \(q_*\)
crosses the cut. The resulting flattening has rank \(\prod_iq_i/q_*\),
and the same cut works for every tensor power.
\end{proof}

In other words, the uniform result controls the nonuniform case. In
logarithmic coordinates, raising the edge weights by amounts
\(\delta_i\ge0\) above a common base \(t\) raises the upper bound by at
most \(\sum_i\delta_i\). The comparison with flattenings is also sharp.
For an odd cycle with nonnegative weights we have
\(\operatorname{MaxCut}(C_5,w)=\sum_iw_i-\min_iw_i\). If we write
\(h=\log_2\aR_5(T_5(\mathbf q))\), \(t=\min_iw_i\) and
\(f=\operatorname{MaxCut}(C_5,w)\), then
\begin{equation}\label{eq:weighted-maxcut-ratio}
f\le h\le f+\frac{t}{12}\le\frac{49}{48}f,
\end{equation}
where the last step uses \(f\ge4t\). Taking all weights equal and
positive shows that the constant \(49/48\) cannot be lowered.

Graph tensors with different dimensions on different edges are the
tensor network states of quantum many-body physics, where the edge
dimensions are called bond dimensions~\cite{LQY,CLVW}. Which networks
can be converted into which others, and at what cost, is the subject
of the resource theory of tensor networks~\cite{CLSWW}.
Proposition~\ref{prop:weighted-pentagon} gives bounds for the pentagon
that hold for all bond dimensions at once, and the ratio between its
upper and lower bounds is at most \(q_*^{1/12}\), independently of how
unequal the bond dimensions are.

\subsection{Fractional pentagon packings}\label{subsec:packing}

Let \(G=(V,E)\) be a finite simple graph with at least two vertices,
and let \(w_e\ge0\) be weights on its edges. Let \(\mathcal C_5(G)\) be
the set of five-cycles in \(G\), induced or not, and define the
fractional packing number
\[
\nu_5^*(G,w)=
\max\Bigl\{\sum_{C\in\mathcal C_5(G)}a_C\Bigm|
a_C\ge0,\ \sum_{C\ni e}a_C\le w_e\ \text{for all }e\in E\Bigr\}.
\]
We also write
\[
h_{G,\F}(w)=\sup_{\phi\in\Phi_{N,\F}}\sum_{e\in E(G)}w_eP_e^{(\phi)}.
\]
By~\eqref{eq:weighted-spectral}, \(h_{G,\F}(w)=\log_2\aR_{\F}(\WT_G(q))\)
whenever \(w_e=\log_2q_e\).

\begin{proposition}\label{prop:packing}
For every nonnegative weight vector \(w\),
\begin{equation}\label{eq:packing}
h_{G,\F}(w)\le\sum_{e\in E}w_e-\frac{11}{12}\,\nu_5^*(G,w).
\end{equation}
In particular, if \(G\) has a fractional decomposition into
five-cycles, then \(\log_2\aR(\GT_2(G))\le49|E|/60\).
\end{proposition}

\begin{proof}
Fix a spectral point \(\phi\) and write \(p_e=P_e^{(\phi)}\). For a
five-cycle \(C\), the graph tensor on \(C\), padded with
one-dimensional factors at the other vertices, has the same rank and
asymptotic rank as \(T_5(2)\). Since every spectral point is bounded by
the asymptotic rank,
\[
\sum_{e\in C}p_e=\log_2\phi(\GT_2(C))\le\frac{49}{12}.
\]
Given a feasible packing, let \(c_e=w_e-\sum_{C\ni e}a_C\ge0\). Using
\(p_e\le1\) we get
\[
\sum_e w_ep_e
=\sum_C a_C\sum_{e\in C}p_e+\sum_e c_ep_e
\le\frac{49}{12}\sum_Ca_C+\sum_e c_e
=\sum_e w_e-\frac{11}{12}\sum_Ca_C.
\]
Taking the supremum over \(\phi\) and the best packing proves
\eqref{eq:packing}. If every edge is covered with total weight one,
then \(5\sum_Ca_C=|E|\), and this gives the last claim.
\end{proof}

The same argument proves an unconditional metric version
of~\eqref{eq:packing}. If \(P\) is the distance matrix of \(N\) points
in a CAT(0) space and \(J-2P\) is copositive, then
\[
\sum_{e\in E(G)}w_eP_e\le
\sum_e w_e-\frac{11}{12}\,\nu_5^*(G,w).
\]
This holds because every five-point principal submatrix satisfies the
hypotheses of Theorem~\ref{thm:intro-opt}, and testing copositivity at
\(e_i+e_j\) shows that \(P_{ij}\le1\).

As an example, consider the Petersen graph, viewed as the graph on the
two-element subsets of \([5]\) in which disjoint subsets are adjacent.
Every vertex has degree three, adjacent vertices have no common
neighbor, and nonadjacent vertices have exactly one. From this one can
check that each of the \(15\) edges lies in four five-cycles, so by
double counting there are \(12\) five-cycles. Giving each cycle weight
\(1/4\) yields
\[
\log_2\aR(\GT_2(\mathrm{Petersen}))\le\frac{49}{4}.
\]
The four subsets that contain \(1\) span a cut of \(12\) edges, so
flattening gives the lower bound \(12\). We do not know whether
\(49/4\) is the best upper bound for this graph. The packing argument
is in the spirit of the covering method of~\cite{CVZ}.
Proposition~\ref{prop:packing} shows that five-cycles can serve as
building blocks in such arguments. Each unit of packed five-cycles
lowers the trivial bound \(\sum_ew_e\) by exactly \(11/12\), which by
Theorem~\ref{thm:intro-rank} is the largest saving that a single
pentagon allows.

\subsection{Lower bounds for all odd cycles}

The grouping that we used for the five-cycle works for every odd cycle
and gives a definite gap above the maximum-cut bound. Here we need the
full rectangular formula of Theorem~\ref{thm:rect}, and not only its value at
\((3,1,1)\).

\begin{proposition}\label{prop:odd-cycle-lower}
For odd \(m\ge3\),
\[
\log_2\aR(\GT_2(C_m))
\ge m-1+\frac{1}{4(m-2)}.
\]
\end{proposition}

\begin{proof}
Label the cycle \(0,\ldots,m-1\). We put the first \(m-1\) vertices
alternately into two blocks and put vertex \(m-1\) into a third block.
Then \(m-2\) edges run between the two alternating blocks and one edge
runs from each of them to the singleton, so the grouped tensor is
rectangular matrix multiplication of shape \((n^{m-2},n,n)\). Grouping
does not increase rank, so the exponent is at least
\(\omega_{\F}(m-2,1,1)\), which by Theorem~\ref{thm:rect} is equal to
\[
(m-2)+1+\frac{(((m-2)+1)-(m-2))^2}{4(m-2)}
=m-1+\frac{1}{4(m-2)}. \qedhere
\]
\end{proof}

Together with Zhou's upper bound~\cite{Zhou} this gives
\[
m-1+\frac{1}{4(m-2)}
\le\log_2\aR(\GT_2(C_m))
\le m-1+\frac{1}{m+1}.
\]
For \(m=5\) the upper bound of the present paper meets the lower bound,
but for \(m\ge7\) the exact exponent is still unknown.
Flattenings give the lower bound \(m-1\) for every odd cycle~\cite{CVZ},
and Proposition~\ref{prop:odd-cycle-lower} shows that the true exponent
is strictly larger for every odd \(m\).

\subsection{A quantum application to entanglement transformations}

Our last application is to quantum information theory. We think of
the five vertices of the cycle as five parties in a quantum network,
each holding part of a shared quantum state. A basic question in the
theory of multipartite entanglement is how much of one entangled
resource is needed to create another when the parties may only act
locally. Here the resource is the GHZ state, which is the standard
example of genuinely multipartite entanglement, and the target is a
pentagon of EPR pairs, in which each pair of neighboring parties shares
one maximally entangled pair.

Over \(\mathbb C\), a graph tensor is exactly such a network state. It
describes an EPR pair on each edge, with the two halves of each pair
held at the two ends of the edge. The five-party GHZ state of rank
\(r\) is
\[
\mathrm{GHZ}^{(5)}_r=\sum_{a=1}^{r}e_a^{\otimes5}.
\]
We ignore nonzero scalar factors, which is natural for stochastic local
operations with postselection. The correspondence between tensor rank
and GHZ cost is standard~\cite{CVZ,VC}, and for stochastic local
operations and classical communication in general we refer
to~\cite{DVC}.

\begin{proposition}\label{prop:ghz}
Let \(m_k(n)\) be the smallest number of five-party binary GHZ states
from which \(k\) copies of \(T_5(n)\) can be obtained exactly by local
linear maps with nonzero probability of success. Then
\begin{align*}
m_k(n)&=\bigl\lceil\log_2\R_5(T_5(n)^{\otimes k})\bigr\rceil,\\
\lim_{k\to\infty}\frac{m_k(n)}{k}&=\frac{49}{12}\log_2n.
\end{align*}
In particular, a pentagon of qubit EPR pairs costs \(49/12\) GHZ states
per copy asymptotically.
\end{proposition}

\begin{proof}
A local linear image of \(\mathrm{GHZ}^{(5)}_r\) has rank at most
\(r\). Conversely, a decomposition of rank \(r\) defines local maps
that send the \(a\)th basis vector of each party to the \(a\)th factor
of the corresponding simple tensor. After scaling, each of these maps
is a contraction and hence a valid local Kraus operator, and the target
state occurs with nonzero probability. Since \(m\) binary GHZ states
are locally equivalent to \(\mathrm{GHZ}^{(5)}_{2^m}\), this proves the
formula for \(m_k(n)\). The limit follows from
Proposition~\ref{prop:all-dimensions}.
\end{proof}

In physical terms, five parties who want to share a pentagon of qubit
EPR pairs, and who start from shared GHZ states, must consume about
\(4.08\) GHZ states per pentagon, and this many suffice in the limit of
many copies. The simple cut bound shows that at least \(4\) are
needed, so the true cost is strictly larger than the naive estimate.

Asymptotic conversion rates between multipartite states were first
studied for the GHZ and W states~\cite{VC15}, and the framework was
later extended from stochastic to exact LOCC
transformations~\cite{JV}. For the triangle of EPR pairs the
corresponding rate is governed by the exponent of matrix
multiplication~\cite{CVZ,VC}. Proposition~\ref{prop:ghz} gives the
analogous exact rate for the pentagon, which is the next odd cycle
after the triangle.

The model allows postselection, and the success probability may shrink
as \(k\) grows. Our result says nothing about deterministic LOCC or
about conversion with success probability bounded away from zero.

\section{Future work}\label{sec:discussion}

Several questions remain open. Since our upper bound is purely
spectral, it would be good to have an explicit family of
decompositions of \(T_5(n)\) whose ranks achieve the exponent
\(49/12\). For odd cycles of length at least seven there is still a gap
between the grouping lower bound of
Proposition~\ref{prop:odd-cycle-lower} and the known upper bound, and a
suitable extension of Theorem~\ref{thm:intro-geometry}, or a better
choice of test vector in the copositivity argument, might close it. We
also do not know all the equality cases of
Theorem~\ref{thm:intro-geometry} or of Theorem~\ref{thm:intro-opt}.

We hope that the method used for Theorem~\ref{thm:intro-geometry} will
be useful for other inequalities between sums of distances. Fixing the
edge directions makes the target a convex function of the edge lengths,
and the low rank of the closure equations forces the extremal
configurations into a small number of patterns. In our case only one
pattern needed more than the triangle inequality, and a few classical
four-point inequalities were enough to finish it. The parallelogram law
at the selected vertex turned out to be the most useful of these,
because the two distances from that vertex to the ends of a longest
side carry the largest weight in the inequality. Since CAT(0) spaces
are Ptolemaic~\cite{Ptolemy}, it would be interesting to find a direct
argument in CAT(0) spaces that does not go through majorization.

\section*{Acknowledgments and use of AI tools}
The problem and the original idea for this work are the author's.
GPT-6 Astra assisted with developing the proofs, and Claude Opus 5.5
assisted with the writing. The author has checked all of the
mathematics and takes full responsibility for the content of the paper.

\bibliographystyle{plainnat}
\bibliography{main}

@article{Strassen1988,
  author = {Volker Strassen},
  title = {The asymptotic spectrum of tensors},
  journal = {Journal f{\"u}r die reine und angewandte Mathematik},
  volume = {384},
  pages = {102--152},
  year = {1988},
  doi = {10.1515/crll.1988.384.102}
}

@article{WZ,
  author = {Avi Wigderson and Jeroen Zuiddam},
  title = {Asymptotic spectra: Theory, applications, and extensions},
  journal = {Bulletin of the American Mathematical Society},
  volume = {63},
  pages = {199--331},
  year = {2026},
  doi = {10.1090/bull/1880}
}

@article{CZ,
  author = {Matthias Christandl and Jeroen Zuiddam},
  title = {Tensor surgery and tensor rank},
  journal = {Computational Complexity},
  volume = {28},
  number = {1},
  pages = {27--56},
  year = {2019},
  doi = {10.1007/s00037-018-0164-8},
  eprint = {1606.04085},
  archivePrefix = {arXiv}
}

@article{CVZ,
  author = {Matthias Christandl and P{\'e}ter Vrana and Jeroen Zuiddam},
  title = {Asymptotic tensor rank of graph tensors: beyond matrix multiplication},
  journal = {Computational Complexity},
  volume = {28},
  number = {1},
  pages = {57--111},
  year = {2019},
  doi = {10.1007/s00037-018-0172-8},
  eprint = {1609.07476},
  archivePrefix = {arXiv}
}

@misc{zhou,
      title={Asymptotic Rank of Weighted Graph Tensors via Completely Positive Bounds}, 
      author={Huanyi Zhou},
      year={2026},
      eprint={2610.34273},
      archivePrefix={arXiv},
      primaryClass={cs.CC},
      url={https://arxiv.org/abs/2610.34273}, 
}

@misc{Geometry,
  author = {{OpenAI}},
  title = {Euclidean geometry in the asymptotic spectrum of tensors},
  year = {2026},
  howpublished = {Preprint, November 24, 2026},
  url = {https://github.com/openai/math/tree/3344da59991ed94f9ba711217b29c1d760badfa6/preprints/Euclidean-geometry-in-the-asymptotic-spectrum-of-tensors-November-24-2026},
  note = {Commit 3344da5. The curvature theorem is the main theorem of the file build/curvature.tex}
}

@misc{Rectangle,
  author = {{OpenAI}},
  title = {The matrix multiplication exponent is nine fourths},
  year = {2026},
  howpublished = {Preprint with Lean formalization, November 24, 2026},
  url = {https://github.com/openai/math/blob/3344da59991ed94f9ba711217b29c1d760badfa6/lean/OAI/LinearAlgebra/SpectralBooks/ExactExponent.lean},
  note = {Commit 3344da5. The rectangular formula is the theorem rectangular\_exponent in OAI.LinearAlgebra.SpectralBooks}
}

@misc{Reshetnyak,
  author = {Anton Petrunin},
  title = {Elements of {Alexandrov} geometry},
  howpublished = {Lecture notes},
  url = {https://anton-petrunin.github.io/alexandrov/alexandrov.pdf},
  note = {Lecture 6, Theorem 6.4},
  year={2024}
}

@article{Ptolemy,
  author = {Thomas Foertsch and Alexander Lytchak and Viktor Schroeder},
  title = {Nonpositive curvature and the {Ptolemy} inequality},
  journal = {International Mathematics Research Notices},
  volume = {2007},
  pages = {rnm100},
  year = {2007},
  doi = {10.1093/imrn/rnm100}
}

@article{Strassen1987,
  author = {Volker Strassen},
  title = {Relative bilinear complexity and matrix multiplication},
  journal = {Journal f{\"u}r die reine und angewandte Mathematik},
  volume = {375/376},
  pages = {406--443},
  year = {1987}
}

@article{Strassen1991,
  author = {Volker Strassen},
  title = {Degeneration and complexity of bilinear maps: some asymptotic spectra},
  journal = {Journal f{\"u}r die reine und angewandte Mathematik},
  volume = {413},
  pages = {127--180},
  year = {1991}
}

@book{BCS,
  author = {Peter B{\"u}rgisser and Michael Clausen and M. Amin Shokrollahi},
  title = {Algebraic Complexity Theory},
  series = {Grundlehren der mathematischen Wissenschaften},
  volume = {315},
  publisher = {Springer},
  address = {Berlin},
  year = {1997}
}

@article{CVZuniversal,
  author = {Matthias Christandl and P{\'e}ter Vrana and Jeroen Zuiddam},
  title = {Universal points in the asymptotic spectrum of tensors},
  journal = {Journal of the American Mathematical Society},
  volume = {36},
  number = {1},
  pages = {31--79},
  year = {2023}
}

@article{CW,
  author = {Don Coppersmith and Shmuel Winograd},
  title = {Matrix multiplication via arithmetic progressions},
  journal = {Journal of Symbolic Computation},
  volume = {9},
  number = {3},
  pages = {251--280},
  year = {1990}
}

@inproceedings{LGU,
  author = {Fran{\c{c}}ois {Le Gall} and Florent Urrutia},
  title = {Improved rectangular matrix multiplication using powers of the {Coppersmith}--{Winograd} tensor},
  booktitle = {Proceedings of the 29th Annual ACM-SIAM Symposium on Discrete Algorithms (SODA)},
  pages = {1029--1046},
  year = {2018}
}

@inproceedings{ADVXXZ,
  author = {Josh Alman and Ran Duan and Virginia {Vassilevska Williams} and Yinzhan Xu and Zixuan Xu and Renfei Zhou},
  title = {More asymmetry yields faster matrix multiplication},
  booktitle = {Proceedings of the 36th Annual ACM-SIAM Symposium on Discrete Algorithms (SODA)},
  year = {2025},
  eprint = {2404.16349},
  archivePrefix = {arXiv}
}

@misc{Pass,
      title={Matrix Multiplication from Polynomial Packings: A Constructive and Elementary Proof of $\omega\le9/4$}, 
      author={Rafael Pass},
      year={2026},
      eprint={2610.28627},
      archivePrefix={arXiv},
      primaryClass={cs.CC},
      url={https://arxiv.org/abs/2610.28627}, 
}

@book{BH,
  author = {Martin R. Bridson and Andr{\'e} Haefliger},
  title = {Metric Spaces of Non-Positive Curvature},
  series = {Grundlehren der mathematischen Wissenschaften},
  volume = {319},
  publisher = {Springer},
  address = {Berlin},
  year = {1999}
}

@article{ReshetnyakOrig,
  author = {Yu. G. Reshetnyak},
  title = {Non-expansive maps in a space of curvature no greater than {$K$}},
  journal = {Siberian Mathematical Journal},
  volume = {9},
  pages = {683--689},
  year = {1968}
}

@article{DVC,
  author = {W. D{\"u}r and G. Vidal and J. I. Cirac},
  title = {Three qubits can be entangled in two inequivalent ways},
  journal = {Physical Review A},
  volume = {62},
  pages = {062314},
  year = {2000}
}

@article{VC,
  author = {P{\'e}ter Vrana and Matthias Christandl},
  title = {Entanglement distillation from {Greenberger}--{Horne}--{Zeilinger} shares},
  journal = {Communications in Mathematical Physics},
  volume = {352},
  number = {2},
  pages = {621--627},
  year = {2017}
}

@incollection{Dur,
  author = {Mirjam D{\"u}r},
  title = {Copositive programming -- a survey},
  booktitle = {Recent Advances in Optimization and its Applications in Engineering},
  publisher = {Springer},
  pages = {3--20},
  year = {2010}
}

@article{VC15,
  author = {P{\'e}ter Vrana and Matthias Christandl},
  title = {Asymptotic entanglement transformation between {W} and {GHZ} states},
  journal = {Journal of Mathematical Physics},
  volume = {56},
  pages = {022204},
  year = {2015},
  eprint = {1310.3244},
  archivePrefix = {arXiv}
}

@misc{JV,
      title={The asymptotic spectrum of LOCC transformations}, 
      author={Asger Kjærulff Jensen and Péter Vrana},
      year={2018},
      eprint={1807.05130},
      archivePrefix={arXiv},
      primaryClass={quant-ph},
      url={https://arxiv.org/abs/1807.05130}, 
}

@article{CLVW,
  author = {Matthias Christandl and Angelo Lucia and P{\'e}ter Vrana and Albert H. Werner},
  title = {Tensor network representations from the geometry of entangled states},
  journal = {SciPost Physics},
  volume = {9},
  pages = {042},
  year = {2020},
  doi = {10.21468/SciPostPhys.9.3.042},
  eprint = {1809.08185},
  archivePrefix = {arXiv}
}

@article{CLSWW,
  author = {Matthias Christandl and Vladimir Lysikov and Vincent Steffan and Albert H. Werner and Freek Witteveen},
  title = {The resource theory of tensor networks},
  journal = {Quantum},
  volume = {8},
  pages = {1560},
  year = {2024},
  doi = {10.22331/q-2024-12-11-1560},
  eprint = {2307.07394},
  archivePrefix = {arXiv}
}

@misc{LQY,
      title={On the geometry of tensor network states}, 
      author={J. M. Landsberg and Yang Qi and Ke Ye},
      year={2012},
      eprint={1105.4449},
      archivePrefix={arXiv},
      primaryClass={math.AG},
      url={https://arxiv.org/abs/1105.4449}, 
}

@article{NW,
  author = {Noam Nisan and Avi Wigderson},
  title = {Lower bounds on arithmetic circuits via partial derivatives},
  journal = {Computational Complexity},
  volume = {6},
  number = {3},
  pages = {217--234},
  year = {1996},
  doi = {10.1007/BF01294256}
}

@inproceedings{LST,
  author = {Nutan Limaye and Srikanth Srinivasan and S{\'e}bastien Tavenas},
  title = {Superpolynomial lower bounds against low-depth algebraic circuits},
  booktitle = {Proceedings of the 62nd Annual IEEE Symposium on Foundations of Computer Science (FOCS)},
  year = {2021}
}

\end{document}